# Constraint condition on transformed relation for generalized acoustics

Jin Hu [1] [*], Xiaoning Liu [2], Gengkai Hu [2] [*]

[1]School of Information and Electronics, Beijing Institute of Technology, Beijing 100081, People's Republic of China.

[2]School of Aerospace Engineering, Beijing Institute of Technology, Beijing 100081, People's Republic of China.

[*]Corresponding author: *bithj@bit.edu.cn, hugeng@bit.edu.cn*

**Abstract**: Contrary to transformation optics (TO), there exist many possibilities for transformed relations of material property and field variable in case of transformation acoustics (TA). To investigate the underlining mechanism and develop a general method that can obtain the full transformed relations, an alternative interpretation to the form-invariance is explored. We consider a spatial transformation, with which a physical phenomenon described in an initial space is transformed to a deformed space, and interpret the mapping by local affine transformation point-by-point. Further, we postulate that the transformed material property and field must rebuild the same physical process, and that the energy must be conserved at each point during the transformation. These conditions impose the constraint on the transformed relation for material property and field. By establishing two local Cartesian frames defined uniquely by the spatial transformation, any physical quantity is shown to first experience a rigid rotation and then a stretch operation during the transformation. We show that the constraint conditions are not enough to determine completely the transformed relation for TA, leaving a possibility to define them differently as found in the literature. New acoustic transformations with constant density or modulus are also proposed and verified by constructing a two-dimensional acoustic cloak.

Finally, we show that the transformed relation is uniquely determined for transformation optics, and discuss how this method can be extended to other transformation physics.

**1. Introduction**

For an electromagnetic (EM) phenomenon, transformation method establishes the equivalency between a curved space and material space [1-4]. It provides an efficient way to find material spatial distribution when wave path is prescribed. With the development of electromagnetic metamaterials, many interesting devices have been proposed, including cloaks [3,5], concentrator and rotator [6,7], beam shifter/bender [8,9], and devices for illusion optics [10]. In parallel, with the help of acoustic metamaterials [11-13], acoustic devices have also been designed by the transformation method for acoustics [14-19], including generalized acoustics, which consider acoustic waves in more complex media in addition to classical fluid. Recently, an acoustic cloak has been demonstrated experimentally [20]. The transformation methods for electromagnetic wave and acoustic wave are called transformation optics (TO) and transformation acoustics (TA), respectively, and both are the results of form-invariance of governing equations under an arbitrary transformation. However, unlike TO, which has clear and unique relation between transformed and initial physical quantity (namely, permittivity, permeability, and electromagnetic fields) [21,22], there exist many possible relations between the transformed and initial mass density and bulk modulus, as well as displacement and pressure for TA. For example, Chen and Chan [14] implicitly assumed that the pressure is unchanged during the transformation, and derived the corresponding transformed relations for mass density and bulk modulus. Milton et al. [15] supposed that the displacement has a special transformation, and they consequently derived the transformed relations for the other physical quantities in a generalized acoustic case. Their results are different from those proposed in [14]. Recently, Cummer et al. [16] proposed a new transformed relation for the displacement (or velocity), and stated that the transformed relation for the displacement proposed in [15] is unsuitable for TA. Norris [17] showed that for a given mapping, the transformed relation for TA is not uniquely defined, and he also pointed out that the transformed relation for the displacement proposed in [15] is possible. Thus, it can be observed that different groups have come up with different formulations and there is no cohesive theory that can link them. This ambiguity also raises another question that whether the TO has no unique transformation. Thus, further efforts are needed to elucidate the mechanism implied in the transformation method itself and find a convenient way to obtain the transformed relation for the physical quantities in general transformation, which is the objective of this study. We will examine the transformation method in a more general context, and figure out the constraint condition on the transformed field and material property during a given spatial transformation.

The paper is arranged as follows: In Section 2, an alternative interpretation to the form-invariance based on the physical point of view is explored; subsequently, the transformation method is examined with the help of deformation view in case of a general mapping. By decoupling a rigid rotation and a stretch operation for each spatial element, the constraint conditions for the transformed relation is derived. In Section 3, application of these conditions to TA is presented. We show that the constraint conditions are not sufficient to determine completely the transformed relations, and new transformed relation is also proposed. In Section 4, some discussions on transformed relation of TO and transformation elastodynamics (TE) are presented, followed by conclusions.

## 2. Constraint condition imposed by transformation method

*2.1 Form-invariance of governing equation*

Consider a specific physical phenomenon prescribed in an initial region $\Omega$, which is governed by a system of differential equation *F,* written in a Cartesian frame as

$$F[\mathbf{x}, t, \mathbf{C}(\mathbf{x}), \mathbf{u}(\mathbf{x}, t)] = 0, \quad \mathbf{x} \in \Omega, \tag{1}$$

where $\mathbf{x}$ is the spatial coordinate and *t* is the time; $\mathbf{C}$ and $\mathbf{u}$ represent the involved material property and related physical field, respectively. They are assumed to be continuous and have continuous derivatives within $\Omega$ [23]. The physical phenomenon described by Eq. (1) gives the relationship between $\mathbf{C}$ and $\mathbf{u}$ in every *point* within $\Omega$. For a concrete physical problem, *F*, $\mathbf{C}$ and $\mathbf{u}$ will be specified together with initial and boundary conditions. In this section, we have presented our discussion in a more general form, and derived general constraint conditions for the transformed material property and physical field when form-invariance of the governing equation is prescribed.

An important question about Eq. (1) is: What is the precision of this equation, or, what is its application scope? For example, can Eq. (1) be used for both homogeneous and inhomogeneous materials (i.e., whether $\mathbf{C}$ is constant or changes with location)? A reasonable governing equation must have clear answer for these questions. If Eq. (1) is equally used for both homogeneous and inhomogeneous materials, we consider it as form-invariant. This is a physical interpretation of form-invariance, different from the mathematical one, as we will explain later. In classical physics, Eq. (1) is usually established by the first-order approximation, i.e., the variation of a quantity around a point is expressed by Taylor series expansion with higher-order terms being neglected. In this paper, we mainly consider the equation with first-order approximation, although the proposed method can be extended to higher order approximation cases.

Consider a spatial mapping $\mathbf{x}' = \mathbf{x}'(\mathbf{x})$; the initial space $\Omega$ is transformed into a new space $\Omega'$, called deformed space from continuum mechanics. Generally, the deformed space prescribes the function that one expects to realize by mapping, as illustrated, e.g., by a rotator in Fig. 1. Once the function is specified, we can determine the spatial material distribution (usually the material becomes inhomogeneous) to realize the desired function; this is a typical inverse problem. However, if the governing equation is form-invariant under the mapping, the transformation method states that the spatial material distribution to realize the prescribed function in the deformed space can be given by the transformed expression of the material property in the initial space. To find the transformed material expressions, usually one has to write down the governing equation in a general curvilinear coordinate system. This so-called "change of variable" method usually involves heavy mathematical algebra. In the present study, we have proposed another method to obtain the transformed material expressions. The main idea is that if governing equation (1) is used for *both homogeneous and inhomogeneous* materials in its known scope, or it is physically form-invariant, then there is no need to find the governing equation for the inhomogeneous materials by changing the variable. The only problem that remains to be resolved is to find the transformed relations of Eq. (1) between homogeneous and inhomogeneous materials. As Eq. (1) is established by the first-order approximation, we can also interpret the general mapping by first-order approximation, i.e., interpret the general mapping locally by affine transformation point-by-point and propose an alternative method to establish the transformed relation. During the transformation, the material property and physical field related by Eq. (1) in the initial space are transported to the deformed space. The transformed material and field cannot be chosen freely and should rebuild the same physical mechanism in the deformed space to fulfill the physical form-invariance. If we can establish a local Cartesian frame at each point in the deformed space, which is uniquely determined by the transformation, the governing equation of first-order approximation written in the local Cartesian frame can be form-invariant due to the assumption of local affine transformation. The local affine transformation guarantees the existence of the local Cartesian system at each point in the transformed space [24]. We insist that the local affine transformation does not indicate that the global transformation is also affine.

For the problem to be addressed in this study, we suppose that the material property and physical field will take the following general forms in the deformed space:

$$\mathbf{C}'(\mathbf{x}') = T_C[\mathbf{C}(\mathbf{x})], \quad \mathbf{u}'(\mathbf{x}',t) = T_u[\mathbf{u}(\mathbf{x},t)], \tag{2}$$

where $\mathbf{C}'$ and $\mathbf{u}'$ are the transformed material property and physical field interpreted in the deformed space, respectively. If the governing equation is also written in a local Cartesian frame in the deformed space, the form-invariance of Eq. (1) forces the transformed material property $\mathbf{C}'$ and physical field $\mathbf{u}'$ in the deformed space $\Omega'$ to satisfy at each point,

$$F[\mathbf{x}',t,\mathbf{C}'(\mathbf{x}'),\mathbf{u}'(\mathbf{x}',t)] = 0. \tag{3}$$

Equation (3) also implies that the transformed material property $\mathbf{C}'$ and physical field $\mathbf{u}'$ should be rearranged to rebuild locally the *same* physical process, which, in fact, imposes constraints on the general transformed relation given by Eq. (2). However, they are too general to be useful at this moment. Therefore, in the following, we will establish two local Cartesian frames – one in the initial space $\Omega$ and the other in the deformed space $\Omega'$ – and we can directly compare Eqs. (1) and (3) written in these two local frames and then establish the transformed relations.

*2.2 Local Cartesian frames*

Consider a mapping $\mathbf{x}' = \mathbf{x}'(\mathbf{x})$, which maps every point $\mathbf{x}$ in $\Omega$ unique to $\mathbf{x}'$ in a new space $\Omega'$, as shown in Fig. 2. It is useful to interpret $\mathbf{x}'$ as another Cartesian coordinate superposed on $\mathbf{x}$ [15]; if the mapping is interpreted by successive local affine transformation, then the mapping defines a deformation field on the initial space $\Omega$, characterized by a deformation gradient tensor $\mathbf{A}$ with elements $A_{ij} = \partial x'_i / \partial x_j$. We ignore the translation part in the affine transformation because it does not affect the physical quantities. The tensor $\mathbf{A}$ can be further decomposed uniquely into a positive definite symmetric tensor and an orthogonal tensor [25]:

$$\mathbf{A} = \mathbf{VR} = \mathbf{RU}, \qquad (4)$$

where $\mathbf{R}$ is an orthogonal tensor characterizing the rigid rotation of a point during the transformation, and $\mathbf{V}$ and $\mathbf{U}$ are the positive definite symmetric tensors describing pure stretch operations. We separately define $\lambda_i$ and $\hat{\mathbf{e}}'_i$ as the eigenvalues and the corresponding eigenvectors of $\mathbf{V}$, i.e., $\mathbf{V} = \lambda_1 \hat{\mathbf{e}}'_1 \hat{\mathbf{e}}'_1 + \lambda_2 \hat{\mathbf{e}}'_2 \hat{\mathbf{e}}'_2 + \lambda_3 \hat{\mathbf{e}}'_3 \hat{\mathbf{e}}'_3$; thus, the eigenvectors $\hat{\mathbf{e}}'_i$ form a local Cartesian frame at each point in the deformed space $\Omega'$. We also define $\hat{\mathbf{e}}_i$ by $\hat{\mathbf{e}}'_i = \mathbf{R}\hat{\mathbf{e}}_i$ (in fact, $\hat{\mathbf{e}}_i$ is the eigenvector of $\mathbf{U}$), and $\hat{\mathbf{e}}_i$ also forms a local Cartesian frame in the initial space $\Omega$. $\hat{\mathbf{e}}_i$ and $\hat{\mathbf{e}}'_i$ can be different from the corresponding global frame $\mathbf{e}_i$, as illustrated in Fig. 2. By establishing these two local Cartesian frames $\hat{\mathbf{e}}_i$ and $\hat{\mathbf{e}}'_i$ for the initial and deformed spaces, respectively, we can write down the governing equation and transformed relation before and after the transformation in $\hat{\mathbf{e}}_i$ and $\hat{\mathbf{e}}'_i$, respectively.

Before proceeding further, we will discuss some properties of the established local Cartesian frames. During mapping, an infinitesimal element $d\Omega$ (with the frame $\hat{\mathbf{e}}_i$ attached) will first be rotated with $\mathbf{R}$, and then rescaled by a factor $\lambda_i$ in the $\hat{\mathbf{e}}'_i$ direction, and finally transformed to $d\Omega'$ (as illustrated in Fig. 2). Therefore, during mapping, any physical quantity attached with the element will experience a rigid rotation operation $\mathbf{R}$ that rotates the physical quantity and the attached base $\hat{\mathbf{e}}_i$ in the initial space to $\hat{\mathbf{e}}'_i$ in the deformed space; then, a stretch operation $\mathbf{V}$ rescales the physical quantity accordingly in $\hat{\mathbf{e}}'_i$, as shown in Fig. 2. The rescaling should ensure the physical quantity to rebuild the same physical mechanism as that in the undeformed space.

To establish the differential relation between these two local Cartesian frames, consider a line element in $d\Omega$, $d\mathbf{x} = dx_i \mathbf{e}_i = d\hat{x}_i \hat{\mathbf{e}}_i$; during mapping (local affine transformation), it is transformed to $d\mathbf{x}'$ in $d\Omega'$ as

$$d\mathbf{x}' = \mathbf{VR}d\mathbf{x} = \lambda_i d\hat{x}_i \hat{\mathbf{e}}'_i = d\hat{x}'_i \hat{\mathbf{e}}'_i, \tag{5}$$

leading to the following differential operation between $\hat{\mathbf{e}}_i$ and $\hat{\mathbf{e}}'_i$:

$$\frac{\partial}{\partial \hat{x}'_i} = \frac{\partial}{\lambda_i \partial \hat{x}_i}. \tag{6}$$

*2.3 Governing equation in the local Cartesian frames*

For the given mapping $\mathbf{x}' = \mathbf{x}'(\mathbf{x})$, we desire to determine the transformed relations for ($\hat{\mathbf{u}}'$, $\hat{\mathbf{u}}$) as well as ($\hat{\mathbf{C}}'$, $\hat{\mathbf{C}}$). In the initial space, the material properties are normalized to be isotropic, and the governing equation is insensitive to the frame direction and can be written in $\hat{\mathbf{e}}_i$ as

$$F[\hat{\mathbf{x}}, t, \hat{\mathbf{C}}(\hat{\mathbf{x}}), \hat{\mathbf{u}}(\hat{\mathbf{x}}, t)] = 0, \quad \text{in } d\Omega \tag{7}$$

As we know that $\hat{\mathbf{C}}$ and $\hat{\mathbf{u}}$ will first experience a rigid rotation $\mathbf{R}$ and then pure stretch operation along the eigenvectors of $\mathbf{V}$ to reach $\hat{\mathbf{C}}'$ and $\hat{\mathbf{u}}'$, we can symbolically write

$$\mathbf{V}_\mathbf{C}\mathbf{R} : \hat{\mathbf{C}} \mapsto \hat{\mathbf{C}}', \qquad \mathbf{V}_\mathbf{u}\mathbf{R} : \hat{\mathbf{u}} \mapsto \hat{\mathbf{u}}'. \tag{8}$$

In the established frame $\hat{\mathbf{e}}'_i$, $\mathbf{V_C}$, $\mathbf{V_u}$ have diagonal forms, which will be determined by the form-invariance of the governing equation during the transformation of $d\Omega$ to $d\Omega'$, because locally, both $\hat{\mathbf{e}}_i$ and $\hat{\mathbf{e}}'_i$ are Cartesian frames, and hence, we have

$$F[\hat{\mathbf{x}}',t,\hat{\mathbf{C}}'(\hat{\mathbf{x}}'),\hat{\mathbf{u}}'(\hat{\mathbf{x}}',t)] = 0. \quad \text{in } d\Omega' \tag{9}$$

With the help of Eqs. (8) and (6), we can express Eq. (9) from the frame $\hat{\mathbf{e}}'_i$ to the frame $\hat{\mathbf{e}}_i$, and compare directly with Eq. (7) to determine $\mathbf{V_C}$, $\mathbf{V_u}$.

It is worth to again point out the difference between the present theory and the usual interpretation for the form-invariance of the transformation method. The form-invariance of a governing equation usually denotes that its form is invariant in any arbitrary *curvilinear* coordinate system [15]. Therefore, to obtain the transformed relations, they should first assume certain transformed relations for some physical quantities, and then express the others in a general arbitrary *curvilinear* coordinate system according to the coordinate transformation laws; these usually need differential geometry technique and heavy calculus, and the results depend on the pre-assumed transformed relations. This interpretation of form-invariance is a mathematical one. However, in the present theory, we emphasize that the local physical mechanism should remain the same during the transformation; as Eq. (1) is equally used for *both homogeneous and inhomogeneous* materials, the form-invariance should be interpreted from the physical point of view. We state that the governing equation only needs to retain its form in local Cartesian frames at each element before and after the transformation, because the governing equation is established itself with a first-order approximation. This form-invariance can be assured by the local affine transformation.

*2.4 Energy conservation condition*

As the mapping just transports a physical mechanism from an initial space to a deformed space, no real physical process manifests during the transformation, and hence, we can assume that at each element, there is no creation or loss of energy during the transformation; in addition, there is no interchange between the different types of energy. If the energy density is denoted by $w = w(\mathbf{C},\mathbf{u})$ in the initial space, then energy conservation leads to $wd\Omega = w'd\Omega'$, where $w'$ is the energy density in the deformed space. With the help of the relation $d\Omega' = \lambda_1\lambda_2\lambda_3 d\Omega$, we have

$$w(\hat{\mathbf{C}},\hat{\mathbf{u}}) = w'(\hat{\mathbf{C}}',\hat{\mathbf{u}}')\lambda_1\lambda_2\lambda_3. \qquad (10)$$

The energy conservation will provide other constraint condition for $\mathbf{V_C}$ and $\mathbf{V_u}$.

In the following, the general idea proposed earlier will be applied to generalized acoustics to determine the transformed relations for the material property and physical field.

**3. Application to generalized acoustic transformation**

*3.1 Constraint conditions for generalized acoustic transformation*

We consider a generalized acoustic wave equation in the context of pentamode material (PM) [17,26],

$$\begin{aligned}\nabla\cdot\boldsymbol{\sigma} &= \boldsymbol{\rho}\cdot\ddot{\mathbf{u}}, \\ \boldsymbol{\sigma} &= \kappa\operatorname{tr}(\mathbf{S}\nabla\mathbf{u})\mathbf{S},\end{aligned} \qquad (11)$$

where $\mathbf{S}$ is a general second-order tensor, $\mathbf{u}$ denotes the displacement vector, $\boldsymbol{\sigma}$ is the stress tensor, $\kappa$ is the bulk modulus, and the density $\boldsymbol{\rho}$ is assumed to have a tensor form [27]. For PM, the corresponding material tensor $\mathbf{C} = \kappa\mathbf{S}\otimes\mathbf{S}$ can be realized at least theoretically [26]. When $\boldsymbol{\rho} = \rho\mathbf{I}$, $\mathbf{S} = \mathbf{I}$, and $\boldsymbol{\sigma} = p\mathbf{I}$, the following classical acoustic wave equation can be recovered:

$$\begin{aligned}\nabla p &= \rho\ddot{\mathbf{u}}, \\ p &= \kappa\nabla\cdot\mathbf{u},\end{aligned} \qquad (12)$$

where $p$ is the acoustic pressure.

Now, we will write down the governing equation (Eq. (11)) in the local Cartesian frames $\hat{\mathbf{e}}_i$ and $\hat{\mathbf{e}}'_i$, respectively. In the initial space, the acoustic wave equation is supposed to have a classical form given by Eq. (12). During the transformation, the material properties $\hat{\rho}\hat{\mathbf{I}}$, $\hat{\kappa}$ and physical fields $\hat{p}\hat{\mathbf{I}}$, $\hat{\mathbf{u}}$ are transformed to $\hat{\boldsymbol{\rho}}'$, $\hat{\kappa}'$ and $\hat{\boldsymbol{\sigma}}'$, $\hat{\mathbf{u}}'$, respectively, and the tensor $\hat{\mathbf{S}} = \hat{\mathbf{I}}$ is transformed to $\hat{\mathbf{S}}'$. According to the previous analysis, there are two operations on each quantity during the transformation: first, a rotation operation from $\hat{\mathbf{e}}_i$ to $\hat{\mathbf{e}}'_i$ by $\mathbf{R}$, then a pure stretch operation in $\hat{\mathbf{e}}'_i$. Taking the displacement as an example, during the transformation, we have: $\hat{\mathbf{u}}' = \mathbf{V}_u \mathbf{R}\hat{\mathbf{u}}$. As the frame $\hat{\mathbf{e}}'_i$ is specially established, which is the principle frame of the stretch $\mathbf{V}$, the pure stretch operation $\mathbf{V}_u$ on the displacement has a diagonal form in this principle frame, and we note it in $\hat{\mathbf{e}}'_i$ by $\mathbf{V}_u = \text{diag}[f_1, f_2, f_3]$. For the rigid rotation operation, the vector $\hat{\mathbf{u}}$, together with the frame $\hat{\mathbf{e}}_i$, are rotated to the new local frame $\hat{\mathbf{e}}'_i$. Therefore, in the local frame $\hat{\mathbf{e}}'_i$, we still have $\mathbf{R}\hat{\mathbf{u}} = [\hat{u}_1, \hat{u}_2, \hat{u}_3]^T$. Finally, the transformed displacement in the frame $\hat{\mathbf{e}}'_i$ can be written as $\hat{\mathbf{u}}' = [f_1\hat{u}_1, f_2\hat{u}_2, f_3\hat{u}_3]^T$. The same idea is applied for the other quantities, and finally, we have

$$\hat{\mathbf{S}}' = \text{diag}[d_1, d_2, d_3],$$
$$\hat{\boldsymbol{\sigma}}' = \hat{p}\,\text{diag}[e_1, e_2, e_3],$$
$$\hat{\mathbf{u}}' = [f_1\hat{u}_1, f_2\hat{u}_2, f_3\hat{u}_3]^T, \qquad (13)$$
$$\hat{\boldsymbol{\rho}}' = \hat{\rho}\,\text{diag}[g_1, g_2, g_3],$$
$$\hat{\kappa}' = \hat{\kappa}h,$$

where $d_i, e_i, f_i, g_i$, and $h$ are the scaling factors on the material property and physical field; they are constant on each transformed element as a result of local affine transformation, and will be determined by the form-invariance of the governing equation. To this end, we write Eq. (12) in $\hat{\mathbf{e}}_i$ as

$$\frac{\partial \hat{p}}{\partial \hat{x}_i} = \hat{\rho} \ddot{\hat{u}}_i,$$

$$\hat{p} = \hat{\kappa}\left(\frac{\partial \hat{u}_1}{\partial \hat{x}_1} + \frac{\partial \hat{u}_2}{\partial \hat{x}_2} + \frac{\partial \hat{u}_3}{\partial \hat{x}_3}\right). \tag{14}$$

After the mapping, the form-invariance of Eq. (11) implies that

$$\nabla \cdot \hat{\boldsymbol{\sigma}}' = \hat{\boldsymbol{\rho}}' \ddot{\hat{\mathbf{u}}}',$$

$$\hat{\boldsymbol{\sigma}}' = \hat{\kappa}' \mathrm{tr}\left(\hat{\mathbf{S}}' \nabla \hat{\mathbf{u}}'\right) \hat{\mathbf{S}}'. \tag{15}$$

As in frame $\hat{\mathbf{e}}'_i$, the second-order tensors have diagonal forms, Eq. (15) can be written in index form without summation as

$$\frac{\partial \hat{\sigma}'_{ii}}{\partial \hat{x}'_i} = \hat{\rho}'_{ii} \ddot{\hat{u}}'_i$$

$$\hat{\sigma}'_{ii} = \hat{\kappa}'\left(\hat{S}'_{11}\frac{\partial \hat{u}'_1}{\partial \hat{x}'_1} + \hat{S}'_{22}\frac{\partial \hat{u}'_2}{\partial \hat{x}'_2} + \hat{S}'_{33}\frac{\partial \hat{u}'_3}{\partial \hat{x}'_3}\right)\hat{S}'_{ii} \tag{16}$$

Using Eqs. (6) and (13), Eq. (16) can be further written as

$$\frac{e_i}{\lambda_i}\frac{\partial \hat{p}}{\partial \hat{x}_i} = g_i f_i \hat{\rho} \ddot{\hat{u}}_i$$

$$e_i \hat{p} = h\hat{\kappa}\left(\frac{d_1 f_1}{\lambda_1}\frac{\partial \hat{u}_1}{\partial \hat{x}_1} + \frac{d_2 f_2}{\lambda_2}\frac{\partial \hat{u}_2}{\partial \hat{x}_2} + \frac{d_3 f_3}{\lambda_3}\frac{\partial \hat{u}_3}{\partial \hat{x}_3}\right)d_i \tag{17}$$

To derive Eq. (17), the following property is used: e.g., $d\hat{u}'_i = f_i d\hat{u}_i$ is employed for the increment of displacement. This is a natural consequence of the local affine transformation, because the scaling factors are constants in any infinitesimal element. By comparing Eq. (17) directly with Eq. (14), the following constraint conditions can be derived:

$$\frac{d_1 f_1}{\lambda_1} = \frac{d_2 f_2}{\lambda_2} = \frac{d_3 f_3}{\lambda_3}, \tag{18a}$$

$$\frac{h d_i^2}{g_i} = \lambda_i^2. \tag{18b}$$

The conservations for strain potential energy and kinetic energy lead to

$$\sum_{i=1}^{3} \hat{\sigma}'_{ii} \frac{\partial \hat{u}'_i}{\partial \hat{x}'_i} = \hat{p} \sum_{i=1}^{3} \frac{e_i f_i}{\lambda_i} \frac{\partial \hat{u}_i}{\partial \hat{x}_i} = \frac{\hat{p}}{\lambda_1 \lambda_2 \lambda_3} \sum_{i=1}^{3} \frac{\partial \hat{u}_i}{\partial \hat{x}_i}, \tag{18c}$$

$$\sum_{i=1}^{3} \hat{\rho}'_{ii} \dot{\hat{u}}_i'^2 = \hat{\rho} \sum_{i=1}^{3} g_i f_i^2 \dot{\hat{u}}_i^2 = \frac{\hat{\rho}}{\lambda_1 \lambda_2 \lambda_3} \sum_{i=1}^{3} \dot{\hat{u}}_i^2. \tag{18d}$$

These then complement the following two additional constraint conditions:

$$\begin{aligned} e_i f_i &= \frac{\lambda_i}{\lambda_1 \lambda_2 \lambda_3}, \\ g_i f_i^2 &= \frac{1}{\lambda_1 \lambda_2 \lambda_3}. \end{aligned} \tag{18e}$$

Totally, we have 11 equations for the 13 unknown scaling variables $d_i$, $e_i$, $f_i$, $g_i$, and $h$, and hence, there is no unique solution; thus, we have some degrees of freedom to choose the transformed relations differently. To illustrate this possibility, in the following, we will give some examples. It can also be noted that $d_i / e_i = \sqrt{\lambda_1 \lambda_2 \lambda_3 / h}$ in Eq. (18) indicates that the pentamode material model is always kept during the transformation.

*3.2 Acoustic transformation with constant pressure*

Let the pressure $p$ be unchanged during the transformation, i.e., $\hat{\boldsymbol{\sigma}}' = \hat{p} \hat{\mathbf{I}}'$ and $\hat{\mathbf{S}}' = \hat{\mathbf{I}}'$; therefore, we set $e_i = 1$ and $d_i = 1$, and then, from Eq. (18), the following unique solution for the scaling factors can be found

$$f_i = \frac{1}{\lambda_j \lambda_k}, \quad g_i = \frac{\lambda_j \lambda_k}{\lambda_i}, \quad h = \lambda_1 \lambda_2 \lambda_3, \qquad (19a)$$
$$i, j, k = 1, 2, 3; i \neq j, i \neq k, j \neq k.$$

Thus, in the frame $\hat{\mathbf{e}}'_i$, the transformed relations for material property and physical field can be expressed as

$$\hat{p}' = \hat{p},$$
$$\hat{\mathbf{u}}' = \frac{\mathrm{diag}[\lambda_1, \lambda_2, \lambda_3]}{\lambda_1 \lambda_2 \lambda_3}[\hat{u}_1, \hat{u}_2, \hat{u}_2]^\mathrm{T},$$
$$\hat{\boldsymbol{\rho}}' = \hat{\rho}\,\mathrm{diag}[\frac{\lambda_2 \lambda_3}{\lambda_1}, \frac{\lambda_1 \lambda_3}{\lambda_2}, \frac{\lambda_2 \lambda_1}{\lambda_3}], \qquad (19b)$$
$$\hat{\kappa}' = \hat{\kappa}\lambda_1 \lambda_2 \lambda_3,$$

or written in a tensor form in a global system due to objectivity of a tensor as

$$p' = p,$$
$$\mathbf{u}' = \frac{\mathbf{VRu}}{\lambda_1 \lambda_2 \lambda_3} = \frac{\mathbf{Au}}{\det \mathbf{A}},$$
$$\boldsymbol{\rho}' = \rho \frac{\lambda_1 \lambda_2 \lambda_3}{\mathbf{V}^2} = \rho \frac{\det \mathbf{A}}{\mathbf{A}\mathbf{A}^\mathrm{T}}, \qquad (19c)$$
$$\kappa' = \kappa \det \mathbf{A}.$$

The transformed relations for $p'$, $\boldsymbol{\rho}'$, and $\kappa'$ agree with those obtained in [14,18], and that of $\mathbf{u}'$ agree with the recent result obtained in [16] in an orthogonal system.

*3.3 Acoustic transformation with unstretched displacement*

We now let the displacement remain unstretched, i.e., $\hat{u}'_i = \hat{u}_i$ or $f_i = 1$, and set $\hat{\mathbf{S}}' = \mathbf{V}/\det \mathbf{A}$ as proposed in [17], i.e., $d_1 = 1/(\lambda_2 \lambda_3), d_2 = 1/(\lambda_3 \lambda_1), d_3 = 1/(\lambda_1 \lambda_2)$; subsequently, from Eq. (18), we can get the unique solution as

$$e_i = \frac{1}{\lambda_j \lambda_k}, \quad g_i = \frac{1}{\lambda_1 \lambda_2 \lambda_3}, \quad h = \lambda_1 \lambda_2 \lambda_3, \qquad (20a)$$
$$i, j, k = 1, 2, 3; i \neq j, i \neq k, j \neq k.$$

The corresponding transformed relations in the global frame are given by

$$\boldsymbol{\sigma}' = \frac{\mathbf{A}p}{\det \mathbf{A}}, \quad \mathbf{u}' = \mathbf{R}\mathbf{u}, \quad \rho' = \frac{\rho}{\det \mathbf{A}}, \quad \kappa' = \kappa \det \mathbf{A}. \qquad (20b)$$

As $\mathbf{C}' = \kappa' \mathbf{S}' \otimes \mathbf{S}'$, it can be noted that the transformed relations for the modulus and density given by Eqs. (19) and (20) can be derived from each other with the following condition: $[\kappa', \rho'_1, \rho'_2, \rho'_3] \leftrightarrow [1/\rho', 1/C'_{1111}, 1/C'_{2222}, 1/C'_{3333}]$. This result agrees with that proposed by Norris [17]. He also noted that the transformed relations given by Eq. (20) could prevent the mass singularity for acoustic cloaks. However, it should be mentioned that in addition to the pressure, the transformed relation for the displacement is also different in these two cases.

*3.4 Acoustic transformation proposed by Milton et al. [15]*

Milton et al. [15] propose the following transformed relation for displacement: $\mathbf{u}' = (\mathbf{A}^\mathrm{T})^{-1} \mathbf{u}$; in case of elastodynamic wave, this condition implies $\hat{u}'_i = \frac{\hat{u}_i}{\lambda_i}$ or $f_i = \frac{1}{\lambda_i}$, and using $\hat{\mathbf{S}}' = \mathbf{V}^2 / \det \mathbf{A}$ as proposed in [17], i.e., $d_1 = \lambda_1 / (\lambda_2 \lambda_3)$, $d_2 = \lambda_2 / (\lambda_3 \lambda_1)$, $d_3 = \lambda_3 / (\lambda_1 \lambda_2)$, Eq. (18) leads to the following unique solution:

$$e_i = \frac{\lambda_i}{\lambda_j \lambda_k}, \quad g_i = \frac{\lambda_i}{\lambda_j \lambda_k}, \quad h = \lambda_1 \lambda_2 \lambda_3, \qquad (21a)$$
$$i, j, k = 1, 2, 3; i \neq j, i \neq k, j \neq k.$$

The corresponding transformed relations in the global frame are derived as

$$\boldsymbol{\sigma}'_i = p \frac{\mathbf{A}\mathbf{A}^\mathrm{T}}{\det \mathbf{A}}, \quad \mathbf{u}' = (\mathbf{A}^\mathrm{T})^{-1} \mathbf{u}, \quad \boldsymbol{\rho}' = \rho \frac{\mathbf{A}\mathbf{A}^\mathrm{T}}{\det \mathbf{A}}, \quad \kappa' = \kappa \det \mathbf{A}, \qquad (21b)$$

and $\mathbf{C}' = \kappa' \mathbf{S}' \otimes \mathbf{S}' = \dfrac{\kappa \mathbf{V}^2 \otimes \mathbf{V}^2}{\det \mathbf{A}}$. We noted that the density tensor $\boldsymbol{\rho}'$ and the elasticity tensor $\mathbf{C}'$ have the same transformed relations as those presented in [15]. These confirm that the transformed relations given by Milton et al. [15] is also admissible for generalized acoustic transformation based on PM theory.

*3.5 Acoustic transformation with one constant material property*

As discussed in Section 3.1, for acoustic transformation, we have 11 constraint equations for determining 13 scaling variables; thus, we can propose new transformed relation. Let us assume that the density is kept constant during the transformation: $\hat{\rho}'(\mathbf{x}') = \hat{\rho}(\mathbf{x})$ or $g_i = 1$. Let us further assume that $\hat{\kappa}' = \hat{\kappa} \det \mathbf{A}$ or $h = \lambda_1 \lambda_2 \lambda_3$, as proposed in [17]; then, from Eq. (18), the following unique solution can be derived:

$$d_i = \left(\dfrac{\lambda_i}{\lambda_j \lambda_k}\right)^{\frac{1}{2}}, \quad e_i = \left(\dfrac{\lambda_i}{\lambda_j \lambda_k}\right)^{\frac{1}{2}}, \quad f_i = \left(\dfrac{1}{\lambda_1 \lambda_2 \lambda_3}\right)^{\frac{1}{2}}, \qquad (22a)$$
$$i, j, k = 1, 2, 3; i \neq j, i \neq k, j \neq k.$$

The corresponding transformed relations in the global frame are given by

$$\mathbf{S}' = \left(\dfrac{\mathbf{A}\mathbf{A}^{\mathrm{T}}}{\det \mathbf{A}}\right)^{\frac{1}{2}}, \quad \mathbf{p}' = p\left(\dfrac{\mathbf{A}\mathbf{A}^{\mathrm{T}}}{\det \mathbf{A}}\right)^{\frac{1}{2}}, \qquad (22b)$$
$$\mathbf{u}' = (\det \mathbf{A})^{-\frac{1}{2}} \mathbf{R} \mathbf{u}, \quad \rho' = \rho, \quad \kappa' = \kappa \det \mathbf{A}.$$

and $\mathbf{C}' = \kappa' \mathbf{S}' \otimes \mathbf{S}'$; thus, without summation of index, we have

$$C'_{iiii} = \lambda_i^2 \kappa. \qquad (23)$$

Similarly, let the bulk modulus remain unchanged, $\hat{\kappa}'(\mathbf{x}') = \hat{\kappa}(\mathbf{x})$ or $h_i = 1$ and $d_i = 1$; then, the following acoustic transformation with constant modulus can also be obtained:

$$e_i = \left(\dfrac{1}{\lambda_1 \lambda_2 \lambda_3}\right)^{\frac{1}{2}}, \quad f_i = \left(\dfrac{\lambda_i}{\lambda_j \lambda_k}\right)^{\frac{1}{2}}, \quad g_i = \dfrac{1}{\lambda_i^2}, \qquad (24a)$$

or

$$\mathbf{S}' = \mathbf{I}, \quad \kappa' = \kappa, \quad \boldsymbol{\sigma}' = p(\frac{1}{\det \mathbf{A}})^{\frac{1}{2}},$$
$$\mathbf{u}' = (\frac{\mathbf{A}\mathbf{A}^{\mathrm{T}}}{\det \mathbf{A}})^{\frac{1}{2}} \mathbf{R}\mathbf{u}, \quad \boldsymbol{\rho}' = \rho(\frac{1}{\mathbf{A}\mathbf{A}^{\mathrm{T}}}).$$

(24b)

The transformed relations for the modulus and density given by Eqs. (22) and (24) also have the following symmetry: $[1/\rho', 1/C'_{1111}, 1/C'_{2222}, 1/C'_{3333}] \leftrightarrow [\kappa', \rho'_1, \rho'_2, \rho'_3]$. We should point out that among the above-mentioned transformations, only the transformed relations given by Eqs. (19) and (24) keep the transformed medium as a fluid, and the other transformations convert a fluid to a more complex material, called the pentamode materials.

To validate the proposed $\rho$-unchanged transformation, in the following construct, a two-dimensional acoustic cloak with the transformed relations for modulus and density given by Eq. (22) will be used. For a cylindrical cloak, the $\rho$-unchanged transformation given by Eq. (22) requires the principal stretches $\lambda_r$, $\lambda_\theta$, $\lambda_z$ to be unity at the outer boundary to satisfy the displacement and pressure continuity conditions [17]. Usually, the outer boundary is fixed during spatial deformation in constructing cloaks, and hence, $\lambda_\theta$ and $\lambda_z$ naturally become unity at the outer boundary; however, $\lambda_r$ at the outer boundary depends on the transformation. The linear transformation $r' = a + r(b-a)/b$ is not applicable to this $\rho$-unchanged transformation, because $\lambda_r = dr'/dr = (b-a)/b \neq 1$ at the outer boundary, where $a$ and $b$ are the radii of the inner and outer boundary of the cloak, respectively. In the following, the nonlinear transformation $r' = ab^2/[(a-b)r + b^2]$ proposed in [28] will be used at the outer boundary, $\lambda_r(r=b) = (b-a)/a = 1$, if $b = 2a$. From this transformation, we can compute $\mathbf{A}$ everywhere in the cloak, and subsequently, the material properties necessary for realizing this cloak are given by Eqs. (22) and (23).

To validate the proposed cloak, we consider a plane acoustic wave incident on the cloak. For the plane wave, the displacement in Eq. (12) can be expressed by a scalar $u$. Eliminating $p$ in Eq. (12) gives the wave equation for the scalar displacement, i.e., the reduced acoustic equation $\nabla \cdot (\kappa \nabla u) - \rho \ddot{u} = 0$. Thus, the same PDE mode (Helmholtz equations) $\nabla \cdot (\mathbf{c} \nabla p) + ap = 0$ of commercial software COMSOL Multiphysics can be used to demonstrate the cloaking effect for a harmonic wave, where $\mathbf{c}$ is a tensor representing the elasticity tensor, just as the method used in [29]. Here, we set $c'_{ii} = \lambda_i^2 \kappa$ from Eq. (23) and $a' = a$. As COMSOL solver requires Cartesian coordinates, it is necessary to write $\mathbf{c}$ in global Cartesian coordinate by the tensor transformation rule, i.e., $c'_{xx} = c'_{rr} \cos^2 \theta + c'_{\theta\theta} \sin^2 \theta$, $c'_{xy} = c'_{yx} = (c'_{rr} - c'_{\theta\theta}) \sin \theta \cos \theta$, and $c'_{yy} = c'_{rr} \sin^2 \theta + c'_{\theta\theta} \cos^2 \theta$. Figure 3 shows the computational domain for a horizontally incident wave, and $a$=0.2 m, $b$=0.4 m. In the simulation, the background medium is set to be water and $\rho = 1 \times 10^3$, $\kappa = 2.18 \times 10^9$ in SI units, and the wavelength of the incident wave is 0.35 m. The material parameters within the cloak are $\rho_{inc} = \rho/5$ and $\kappa_{inc} = \kappa$. The simulation of the cloak constructed by the proposed $\rho$-unchanged transformation is shown in Fig. 4, and the simulation result confirms the validity of the proposed transformation. The imperfection of the simulated result is believed to be caused by the numerical simulation. When the Helmholtz equation is solved in the simulation, the parameter $\mathbf{c}$ will tend to be infinite with a higher order than the stretch $\lambda_\theta$ near the inner boundary (see Eq. (23)), and more refined discretization is needed with a cost of computation time.

## 4. Discussions and conclusions

One should be aware that the proposed method relies on the physical model of the governing equation, i.e., the precision of the governing equation in describing the physical phenomenon for homogeneous and inhomogeneous materials. The first-order approximation of the governing equation can be equally used for homogeneous and inhomogeneous materials, because the physical model is based on a point. The proposed method can work well for those equations. Let us take the electromagnetic transformation as an example. Maxwell's equations in Cartesian coordinate read

$$\nabla \times \mathbf{E} = -\mathbf{\mu}\dot{\mathbf{H}}, \quad \nabla \times \mathbf{H} = +\mathbf{\varepsilon}\dot{\mathbf{E}}. \tag{25}$$

It is a standard first-order approximation physical model and equally used for both homogeneous and inhomogeneous materials. Maxwell's equations possess a special symmetry, which indicates that the material parameters and fields should have the same transformation, respectively, or more explicitly, $\mathbf{\varepsilon}' = \mathbf{\mu}'$ and $\mathbf{E}' = \mathbf{H}'$, if $\mathbf{\varepsilon} = \mathbf{\mu}$ and $\mathbf{E} = \mathbf{H}$. This condition makes it possible to analyze the electromagnetic transformation only by one of the equations in Eq. (25). The material parameters in the initial region are assumed to be isotropic, i.e., $\mathbf{\varepsilon} = \varepsilon \mathbf{I}$ and $\mathbf{\mu} = \mu \mathbf{I}$. According to the method proposed in Section 2, the transformed relations for the material property and physical field take the following forms in the local Cartesian frame $\hat{\mathbf{e}}'_i$:

$$\begin{aligned}
\hat{\mathbf{\varepsilon}}' &= \hat{\varepsilon}\,\mathrm{diag}[\,a_1, \quad a_2, \quad a_3\,], \\
\hat{\mathbf{\mu}}' &= \hat{\mu}\,\mathrm{diag}[\,a_1, \quad a_2, \quad a_3\,], \\
\hat{\mathbf{E}}' &= [b_1\hat{E}_1 \quad b_2\hat{E}_2 \quad b_3\hat{E}_3]^{\mathrm{T}}, \\
\hat{\mathbf{H}}' &= [b_1\hat{H}_1 \quad b_2\hat{H}_2 \quad b_3\hat{H}_3]^{\mathrm{T}}.
\end{aligned} \tag{26}$$

It is easy to establish the following unique solution according to the method proposed in Section 2:

$$a_i = \frac{\lambda_i}{\lambda_j \lambda_k}, \quad b_i = \frac{1}{\lambda_i}, i, j, k = 1, 2, 3; i \neq j, i \neq k, j \neq k. \tag{27}$$

The corresponding transformed relations in the global frame are derived as

$$\boldsymbol{\varepsilon}' = \frac{\mathbf{A}\boldsymbol{\varepsilon}\mathbf{A}^{\mathrm{T}}}{\det \mathbf{A}}, \quad \boldsymbol{\mu}' = \frac{\mathbf{A}\boldsymbol{\mu}\mathbf{A}^{\mathrm{T}}}{\det \mathbf{A}}, \quad \mathbf{E}' = (\mathbf{A}^{\mathrm{T}})^{-1}\mathbf{E}, \quad \mathbf{H}' = (\mathbf{A}^{\mathrm{T}})^{-1}\mathbf{H}. \tag{28}$$

Equation (28) agrees with the known results in the literatures. It is also shown that the transformed relations for electromagnetic transformation are uniquely determined.

It is also very interesting to examine whether we can obtain transformed relations for elastodynamics by this method. The corresponding governing equation, namely Navier's equation, reads

$$\nabla \cdot \boldsymbol{\sigma} = \rho \frac{\partial^2 \mathbf{u}}{\partial t^2}, \quad \boldsymbol{\sigma} = \mathbf{C} : \nabla \mathbf{u}. \tag{29}$$

The balance equation of angular momentum by first-order approximation is written as (without summation) [25]

$$\sigma_{ij} + \frac{1}{2}\frac{\partial \sigma_{ij}}{\partial x_i}dx_i = \sigma_{ji} + \frac{1}{2}\frac{\partial \sigma_{ji}}{\partial x_j}dx_j, \quad i,j=1,2,3, i \neq j, \tag{30}$$

where $dx_i$ is the size of an element. To avoid mathematic difficulty, Eq. (30) is further simplified to $\sigma_{ij} = \sigma_{ji}$ by *zero*-order approximations, and is usually omitted in Navier's equation because it is included by the symmetry of classical elasticity tensor, i.e., $C_{ijkl} = C_{ijlk} = C_{jikl} = C_{klij}$. Thus, the Navier's equation with symmetric elasticity tensor in fact uses zero-order approximation. Its physical model is based on a finite area, as shown in Eq. (30), rather than a point that cannot support moment. As a result, it cannot be equally used for homogeneous and inhomogeneous materials. If the material perturbations inside the element are considered instead of ignoring them by zero-order approximations, the governing equation should be Willis' equation [30,31]. If the angular momentum balance is totally ignored and thus there are no zero-order approximations, then the asymmetric elasticity tensor can appear [32], which is in fact beyond the framework of classical elastodynamics, and asymmetric elasticity tensor is difficult if not impossible to realize in practice. Therefore, the transformed relations obtained by the proposed method in the context of classical Navier's equation [33, 34] should be limited in high-frequency or slowly varying materials [35], i.e., the application scope of Navier's equation. Recently, Norris et al. [36] also developed a comprehensive theory for elasto-mechanical transformation, and their results are still based on mathematical interpretation of form-invariance.

In conclusion, we have explored the underlining mechanism of transformation method and developed a general method to obtain the full transformed relations by introducing physical interpretation of form-invariance. For a given spatial mapping that will realize a desired function, a physical phenomenon prescribed in the initial space is transported to a deformed space defined by the mapping. The transformed material property and physical field are constrained by the following requirements: (1) They should rebuild the same physical process as prescribed in the initial space; (2) every type of energy on each element is assumed to be conserved during transformation. On interpreting a general mapping by the first-order approximation or local affine transformation point-by-point, any physical quantity will experience a rigid rotation and then a rescaling in amplitude as the spatial element. By establishing two local Cartesian frames, the form-invariance of governing equation in the local Cartesian frames and energy conservation will give the constraint condition for the transformation. The constraint conditions for transformed relations of material property and physical field are then derived for TA and TO, respectively. It is found that the constraint conditions are less than the scaling variables for TA, leaving a possibility to define transformed relations differently; this also explains the different acoustic transformations proposed in the literature. New acoustic transformations with constant density or modulus are also proposed and validated by constructing a two-dimensional acoustic cloak. Finally, we have shown that for TO, the transformed relations are uniquely determined, and for TE, the transformed relations should be limited in some scope. The proposed theory provides a general method to determine the transformed relations for any physical process governed by a set of PDE within its application scope, and thus, we have a convenient method to explore the transformation method in a vast range of potential dynamical systems.

.

## Acknowledgments

This work was supported by the National Natural Science Foundation of China (10702006, 10832002 and 11172037), and the National Basic Research Program of China (2006CB601204).

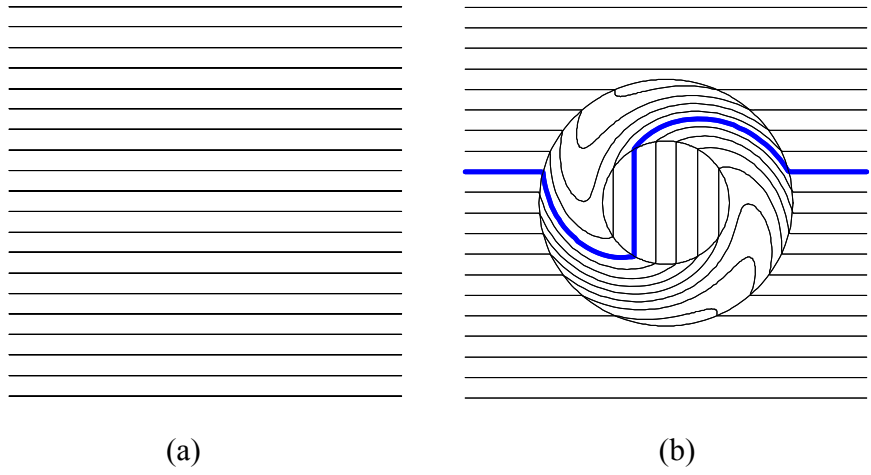

(a)                    (b)

Figure 1. A sketch of transformation method: (a) Initial space; (b) deformed space characterizing the designed function.

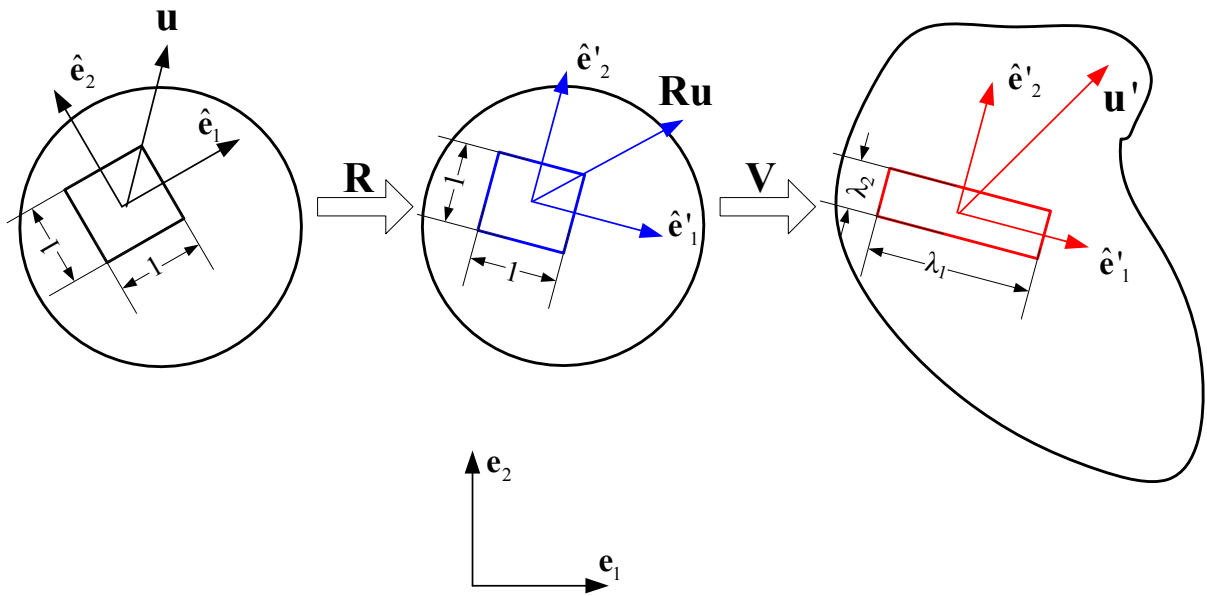

Figure 2. Rigid rotation and stretch operation during a transformation at each point

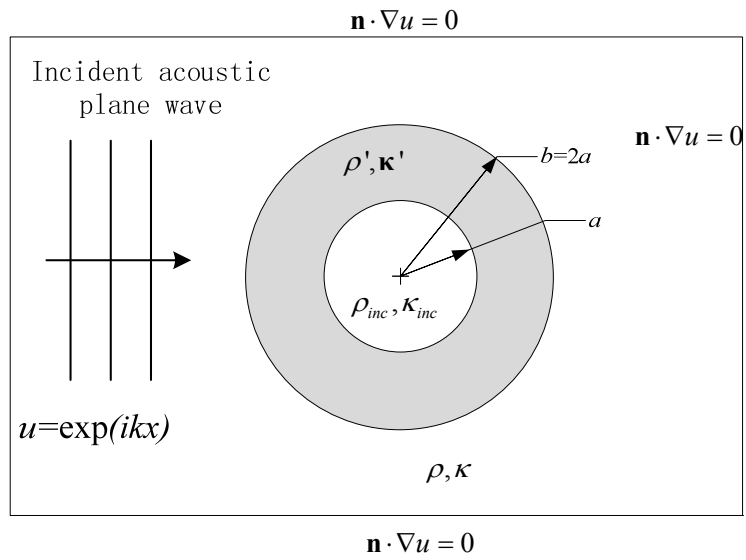

Figure 3. Computational domain and simulation for a left-incident acoustic cloak based on Helmholtz equation.

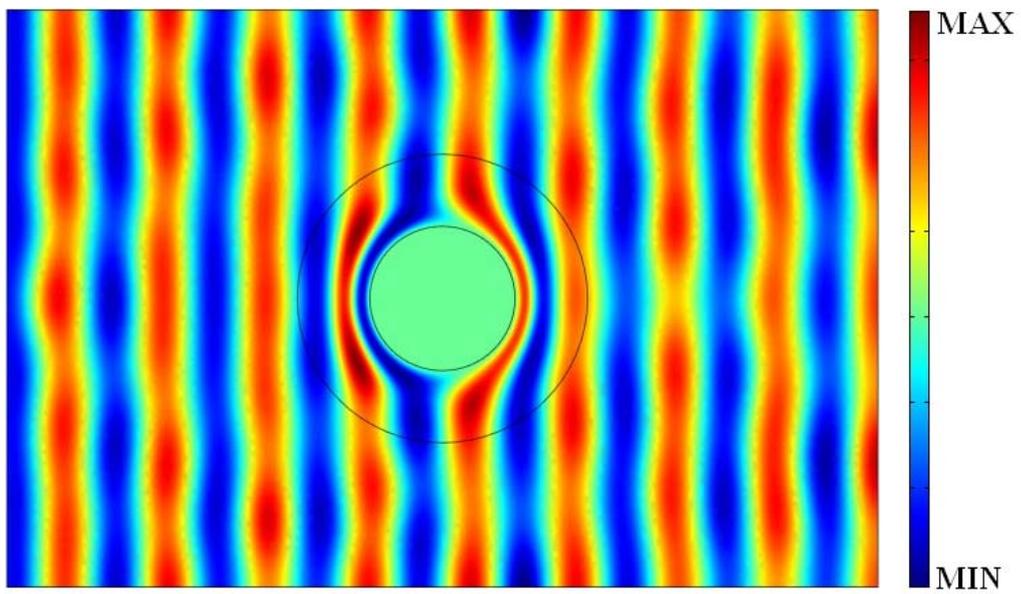

Figure 4. Simulation of the displacement field around the $\rho$-unchanged acoustic cloak.